\documentclass[twocolumn]{aastex61}

\received{May 12, 2017}
\revised{June 18, 2017}
\accepted{June 20, 2017}
\submitjournal{ApJ}

\usepackage{amsmath}
\usepackage{amssymb}
\usepackage{graphicx,graphics}
\usepackage{rotating}
\usepackage{natbib}
\usepackage{color}
\usepackage{comment}
\usepackage{apjfonts}

\usepackage{listings}
\usepackage{hyperref}
\hypersetup
{
    pdfauthor={M.~Kokubo et al.},
    pdftitle={Kokubo et al. 2017},
}

\shorttitle{AO-assisted optical IFU observation for  the FRB~121102 host galaxy}
\shortauthors{M.~Kokubo et al.}

\begin{document}

\title{H$\alpha$ intensity map of the repeating fast radio burst FRB~121102 host galaxy from Subaru/Kyoto~3DII AO-assisted optical integral-field spectroscopy\footnote{Based on data collected at Subaru Telescope, which is operated by the National Astronomical Observatory of Japan.}}

\correspondingauthor{Mitsuru Kokubo}
\email{mkokubo@astr.tohoku.ac.jp}

\author{Mitsuru Kokubo}
\altaffiliation{JSPS Fellow}
\affil{Astronomical Institute, Tohoku University, 6-3 Aramaki, Aoba-ku, Sendai, Miyagi 980-8578, Japan}

\author{Kazuma Mitsuda}
\affil{Department of Astronomy, School of Science, the University of Tokyo, 7-3-1 Hongo, Bunkyo-ku, Tokyo 113-0033, Japan}
\affil{Institute of Astronomy, the University of Tokyo, 2-21-1 Osawa, Mitaka, Tokyo 181-0015, Japan}

\author{Hajime Sugai}
\affil{Kavli Institute for the Physics and Mathematics of the Universe (WPI), Institutes for Advanced Study, University of Tokyo, Kashiwa, Chiba 277-8583, Japan}

\author{Shinobu Ozaki}
\affil{National Astronomical Observatory of Japan, 2-21-1 Osawa, Mitaka, Tokyo 181-8588, Japan}

\author{Yosuke Minowa}
\affil{Subaru Telescope, National Astronomical Observatory of Japan, 650 North A'hoku Place, Hilo, HI, 96720, USA}

\author{Takashi Hattori}
\affil{Subaru Telescope, National Astronomical Observatory of Japan, 650 North A'hoku Place, Hilo, HI, 96720, USA}

\author{Yutaka  Hayano}
\affil{National Astronomical Observatory of Japan, 2-21-1 Osawa, Mitaka, Tokyo 181-8588, Japan}

\author{Kazuya Matsubayashi}
\affil{National Astronomical Observatory of Japan, 2-21-1 Osawa, Mitaka, Tokyo 181-8588, Japan}

\author{Atsushi Shimono}
\affil{Kavli Institute for the Physics and Mathematics of the Universe (WPI), Institutes for Advanced Study, University of Tokyo, Kashiwa, Chiba 277-8583, Japan}

\author{Shigeyuki Sako}
\affil{Institute of Astronomy, the University of Tokyo, 2-21-1 Osawa, Mitaka, Tokyo 181-0015, Japan}

\author{Mamoru Doi}
\affil{Institute of Astronomy, the University of Tokyo, 2-21-1 Osawa, Mitaka, Tokyo 181-0015, Japan}
\affil{Research Center for the Early Universe, Graduate School of Science, The University of Tokyo, Hongo, 7-3-1, Bunkyo-ku, Tokyo, 113-0033, Japan}



\begin{abstract}

We present the H$\alpha$ intensity map of the host galaxy of the repeating fast radio burst FRB~121102 at a redshift of $z=0.193$ obtained with the AO-assisted Kyoto~3DII optical integral-field unit mounted on the 8.2-m Subaru Telescope.
We detected a compact H$\alpha$-emitting (i.e., star-forming) region in the galaxy, which has a much smaller angular size [$ < 0''.57$ (1.9~kpc) at full width at half maximum (FWHM)] than the extended stellar continuum emission region determined by the Gemini/GMOS $z'$-band image [$\simeq 1''.4$ (4.6~kpc) at FWHM with ellipticity $b/a=0.45$].
The spatial offset between the centroid of the H$\alpha$ emission region and the position of the radio bursts is $0''.08 \pm 0''.02$ ($0.26 \pm 0.07$~kpc), indicating that FRB~121102 is located within the star-forming region.
This close spatial association of FRB~121102 with the star-forming region is consistent with expectations from young pulsar/magnetar models for FRB~121102, and it also suggests that the observed H$\alpha$ emission region can make a major dispersion measure (DM) contribution to the host galaxy DM component of FRB~121102.
Nevertheless, the largest possible value of the DM contribution from the H$\alpha$ emission region inferred from our observations still requires a significant amount of ionized baryons in intergalactic medium (the so-called `missing' baryons) as the DM source of FRB~121102, and we obtain a 90\% confidence level lower
limit on the cosmic baryon density in the intergalactic medium in the low-redshift universe as $\Omega_{\text{IGM}} > 0.012$.

\end{abstract}

\keywords{intergalactic medium, galaxies: dwarf, galaxies: ISM, instrumentation: adaptive optics, stars: magnetars, stars: neutron, techniques: imaging spectroscopy}



\section{Introduction}
\label{sec:intro}

Fast radio bursts (FRBs) are a new class of millisecond duration radio bursts \citep{lor07,tho13,pet16}, characterized by their large dispersion measures (DM), suggesting that FRBs are located at cosmological distances with excess DM due to free electrons in the intergalactic medium (IGM).
However, because FRBs are one-off events, it is generally difficult to identify their host galaxies and thus to determine spectroscopic redshift.

Among the FRBs discovered to date \citep{pet16}\footnote{\href{http://www.astronomy.swin.edu.au/pulsar/frbcat/}{http://www.astronomy.swin.edu.au/pulsar/frbcat/}},  FRB~121102 is unique in terms of its repetitiveness \citep[see also][]{cal17}.
About $30$ bursts over 2012-2016 have been reported in the direction of FRB 121102 at 1.1$-$3.5 GHz with a time-constant DM of $\sim$558~$\text{pc}$~$\text{cm}^{-3}$, and frequency-dependent arrival time $t_a\propto \nu^{m}$ with $m=-2$ as expected for signals propagating through cold plasma \citep{spi14,spi16,sch16,mar17}.
The 12 Arecibo FRB 121102 radio pulses \citep{spi16} show pulse widths spanning 3$-$9 ms.
No scattering tail is observed in the pulses, suggesting that the observed pulse widths represent the intrinsic width of the emission \citep{sch16}.
FRB~121102 shows no temporal evolution of flux and pulse width \citep{spi16,mar17}.
It should be noted that its repetitiveness rules out the possibility that FRB 121102 is related to a catastrophic event such as collapse of a supramassive neutron star to a black hole \citep{fal14} or a binary neutron star merger \citep{tot13}.

Based on multiple detections of the radio bursts in the 2.5$-$3.5 GHz band in the direction of FRB~121102 during an interferometric localization campaign with the Karl G. Jansky Very Large Array, \cite{cha17} reported the average position of the burst source as $\alpha=05^{h}31^{m}58^{s}.70$ and $\delta=+33^{\circ}08'52.5''$ (J2000) with 1$\sigma$ uncertainty of about 100 milliarcseconds (mas).
Then, \cite{mar17} reported an independent localization of FRB~121102 using the European Very Long Baseline Interferometry Network with a better precision of $\lesssim$ 10 mas \citep[see Table~1 of][]{mar17}.
\cite{cha17} also identified a host galaxy candidate at a position consistent with FRB~121102 in optical images taken by the 10-m Keck telescope/LRIS ($R=24.9 \pm 0.1$ AB mag) and 8-m Gemini North telescope/GMOS \citep[$r=25.1 \pm 0.1$ AB mag; see also][]{ten17}.
\cite{ten17} carried out optical spectroscopy covering the wavelength range from 4650 to 8900~\AA for the host galaxy of FRB~121102 with Gemini/GMOS on November 9 and 10, 2016, with a total exposure time of 4.5 hours.
Note that they used a $1''$-width slit, and the host galaxy was not spatially resolved during the spectroscopic observations due to the poor seeing conditions \citep[see ][fir details]{ten17}.
H$\alpha$, H$\beta$, [\ion{O}{3}]$\lambda \lambda$4959,5007 and [\ion{S}{2}]$\lambda\lambda 6717, 6731$ emission lines are detected in the GMOS spectrum, indicating that the host galaxy of FRB~121102 is a low-metallicity ($\log_{10}(\text{[O/H]})+12<8.7$), star-forming dwarf galaxy at a redshift of $z=0.19273 \pm 0.00008$ \citep[][]{ten17,mar17}.
The host galaxy has a Gaussian semi-major axis width of $\sigma_{a}=0''.59$ (1.95~kpc) with ellipticity $b/a=0.45$ (measured on the $z'$-band Gemini/GMOS image), and a stellar mass of $M_{*}\sim(4-7)\times10^{7}M_{\odot}$ \citep{ten17}.
The integrated H$\alpha$ emission line luminosity implies that the star-formation rate of the FRB~121102 host galaxy is $\sim 0.4$ $M_{\odot}$/yr \citep{ten17}.
\cite{ten17} noted that hydrogen-poor superluminous supernovae (SLSNe) and long gamma-ray bursts (LGRBs) are known to reside preferentially in such dwarf galaxies, suggesting that FRB~121102 radio bursts originate from a young neutron star or a magnetar, formed by one of these unusual supernova explosions \citep[see][and references therein]{lun14,lel15,met17,nic17}.
Based on the redshift determination, the burst energies of each of the radio bursts of FRB~121102 can be evaluated as $\sim 10^{38}~\text{erg}~(\delta\Omega/4\pi)(A_{\nu}/0.1~\text{Jy ms}) (\Delta \nu/1~\text{GHz})$, where $A_{\nu}$, $\Delta \nu$, and $\delta\Omega$ are the fluence, bandwidth, and opening angle of the radio bursts, respectively, and the energetics of FRB~121102 may fall within the range expected from the magnetosphere of a young neutron star (pulsar) or a magnetar \citep[][and references therein]{tot13,cha17,ten17,met17}.

To constrain the source models for FRB~121102 further, it is very important to examine the local environment in the host galaxy around the position of FRB~121102.
As noted by \cite{ten17}, the $r'$ and $i'$-band images of the FRB~121102 host galaxy, which include not only the continuum emission but also the strong H$\beta$, [\ion{O}{3}], or H$\alpha$ emission lines, show $\sim 0.1$ arcsec offsets from the continuum-dominated $z'$-band image.
This suggests that the host galaxy has at least one \ion{H}{2} region at a slight offset from the galaxy center \citep{ten17}. 
FRB~121102 may reside in the \ion{H}{2} region, and the observed DM may be partly due to free electrons in this \ion{H}{2} region.
Here, we present integral-field spectroscopic observations obtained with the Kyoto Tridimensional Spectrograph II (Kyoto~3DII) optical integral-field unit (IFU) mounted on the 8.2-m Subaru Telescope to determine the position of the centroid of the H$\alpha$ emission region accurately and constrain its spatial size.
In Section~\ref{observation}, we describe the AO-assisted Subaru/Kyoto~3DII IFU observation for the host galaxy of FRB~121102 in detail.
In Section~\ref{sec:halpha_map}, we present the obtained spectra and the H$\alpha$ intensity map of the host galaxy.
We derive the positional offsets between the centroids of the stellar continuum and the H$\alpha$ emission region of the host galaxy and the position of the FRB~121102 radio bursts, and discuss the possible DM contributions from the H$\alpha$ emission region to FRB~121102 in Section~\ref{discuss}.
Our conclusions are summarized in Section~\ref{conclusion}.

Following \cite{ten17}, throughout this paper we assume the Planck15 cosmology \citep{pla16}; $H_0 = 67.7$ km/s/Mpc, $\Omega_\text{m}=0.307$, and $\Omega_{\text{b}}=0.049$ flat $\Lambda$CDM.
The angular size scale at $z=0.19273$ is 3.31~kpc/$''$ \citep{wri06}.

\section{Observation}
\label{observation}

\begin{figure}[tbp]
\center{
\includegraphics[clip, width=3.2in]{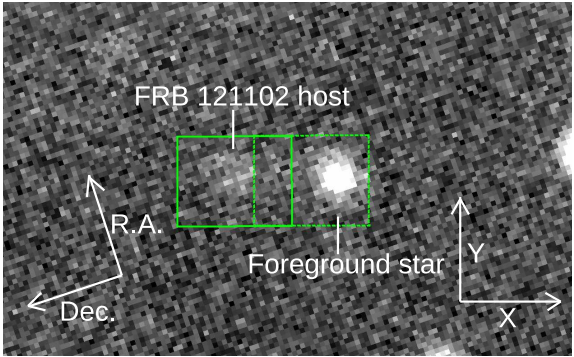}
}
 \caption{Keck/LRIS optical $R$-band image ($0''.135$/pixel) of the host galaxy of FRB~121102 \citep[$R=24.9 \pm 0.1$ AB mag;][]{cha17}. Solid and dashed squares denote the Subaru/Kyoto~3DII $3''.21 \times 2''.52$ FoV locations in which the integral-field spectra of the host galaxy and the no-grism image of the $i=22.7$ AB mag foreground star are obtained, respectively.
 }
 \label{fig:kecklris}
\end{figure}

\begin{figure*}[tbp]
\vspace{-1cm}
\center{
\includegraphics[clip, width=8.2in]{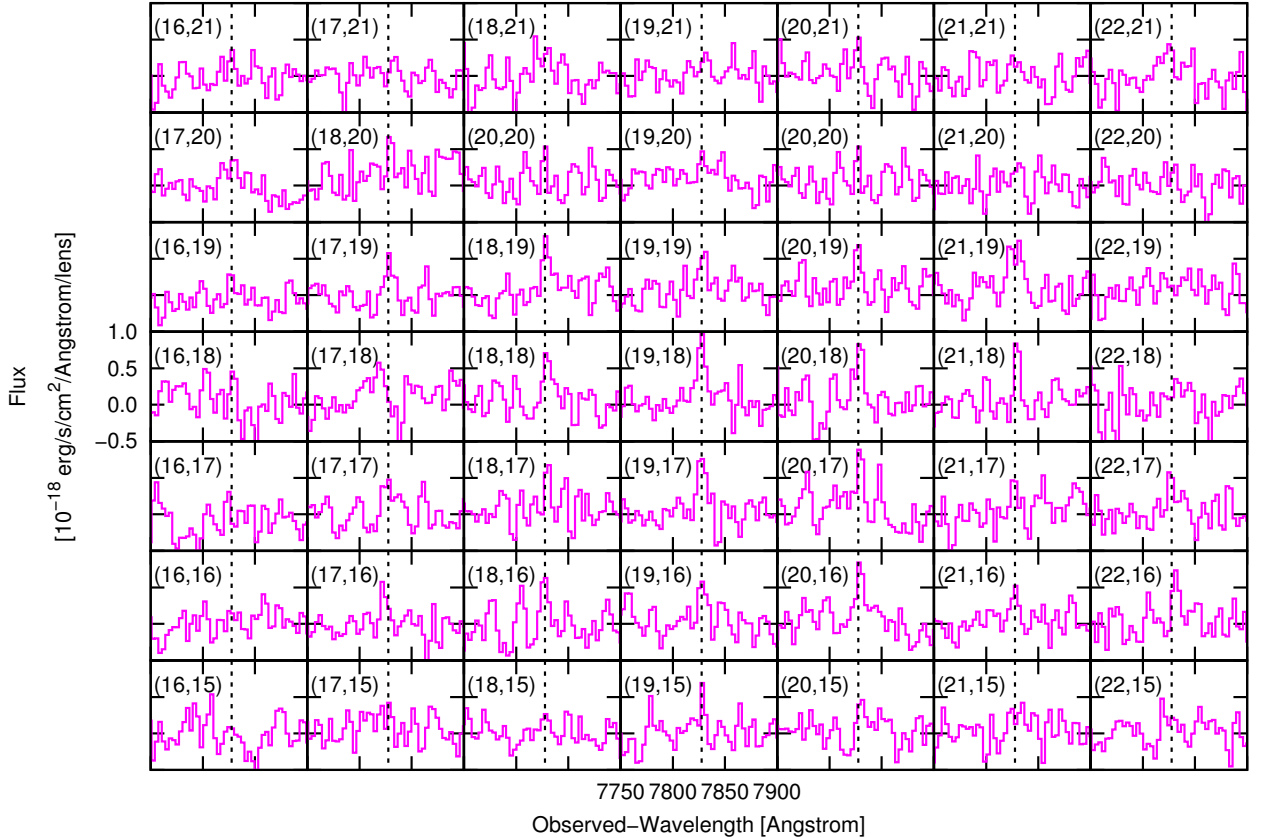}
}
\vspace{-2cm}
 \caption{
 Kyoto~3DII IFU spectra at spaxels around the spaxel with peak H$\alpha$ intensity $(X,Y)=(19,18)$ at wavelengths around the rest-frame H$\alpha$ emission (1 lens = 1 spaxel = $0''.0868 \times 0''.0868$). Galactic extinction is corrected.
 Dotted lines indicate the observed-frame wavelength of H$\alpha$ at a redshift of $z=0.19273$ \citep{ten17}.
 }
 \label{fig:each_spaxel}
\end{figure*}

We observed the host galaxy of FRB~121102 on February 9, 2017, with Kyoto~3D-II optical IFU \citep[][]{sug10,mat16} at a Nasmyth focus of the Subaru 8.2-m telescope on Mauna Kea, Hawaii \citep{iye04}.
Kyoto~3D-II was used in combination with a 188-element adaptive optics (AO) system Subaru AO188 \citep{hay08,hay10,min10}, enabling AO-assisted optical integral-field spectroscopy at wavelengths of $\gtrsim$ 6400 \AA\ \citep{mat16}.
In micro-lens array IFU mode, the field of view (FoV) of Kyoto~3D-II is $3''.21 \times 2''.52$, where there are 37$\times$29 lenses (corresponding to spatial pixels or spaxels) in the lenslet array for object observation (the spatial sampling rate is $0''.0868 \pm 0''.0002$ lens${}^{-1}$).\footnote{The lens (pixel) scale is accurately evaluated from Subaru/Kyoto~3DII integral-field spectroscopy data of a quadruple quasar H1413+117 obtained during the same Subaru/Kyoto~3DII observation run as the FRB~121102 host galaxy.
Uncertainty of the instrumental position angle of $\pm0.16$ deg (described later) is also evaluated from the same data.}
Throughout this paper, we use the characters (X, Y) to denote the coordinates of the $37 \times 29$ spaxels.
Kyoto~3D-II is able to obtain sky spectra with the object spectra with FoV of $3''.21 \times 0''.61$ simultaneously, where the sky regions are separated by $\sim 29''$ from the object regions.
We used a newly installed red-sensitive Hamamatsu fully depleted charge-coupled device \citep[CCD;][]{mit16} and an order-sort filter \citep[referred to as ``No.5'' in][]{mat16}, covering the wavelength range from 7300 to 9200 \AA\ with a wavelength sampling rate of 3.8~\AA/pixel \footnote{The wavelength sampling rate of the new CCD (3.8~\AA/pixel) is slightly different from the previous CCD \citep[3.3~\AA/pixel; Table~5 of][]{mat16}.}.
The size of the micropupil images on the CCD measured at the full width at half maximum (FWHM) is 1.69 pixels \citep{mit16}, and thus the instrumental spectral resolution is $\sim$ 6.42 \AA\ at FWHM ($2.73$~\AA\ at 1$\sigma$).
An atmospheric dispersion corrector unit was used during our observations.

We used the laser guide star AO mode of AO188 \citep{hay08,hay10}.
A bright field star PSO J053200.260+330833.843 ($\alpha=$05$^{h}$32$^{m}$00$^{s}$.260, $\delta=$+33${}^{\circ}$08'33''.80), separated by $\sim27''$ from the location of FRB~121102, is selected as a tip-tilt AO guide star.
Although the AO-assisted Subaru/Kyoto~3DII optical IFU has the potential to achieve a FWHM point source image of $\sim 0''.2$ under good natural seeing conditions at Mauna Kea Observatory \citep{mat16}, the poor natural seeing conditions ($\sim 1''.2-1''.4$) on the observing night on February 9, 2017, resulted in an AO-corrected FWHM of $\sim 0''.5-0''.6$ (see Section~\ref{sec:halpha_map}).

The on-sky position angle ($PA$)\footnote{Note that the Kyoto~3DII pixel (X,Y) coordinate is inverted in the X direction from the sky coordinate.} of the FoV of Kyoto~3DII was set to $108.99 \pm 0.16$ deg to align the host galaxy of FRB~121102 horizontally with a nearby $i=22.7$~AB~mag foreground star ($2''.8$ away from FRB 121102's position; Figure~\ref{fig:kecklris}) \citep[see][]{ten17}.
The total on-source exposure time is 3 hours (9 exposures $\times$ 1200 sec) with no dithering.
Moreover, to calibrate the relative sky coordinates from the $2''.8$-away foreground star to the FRB 121102 host, a no-grism (7300-9200~\AA) image for the foreground star with an exposure time of 300~s was obtained by applying $\text{X}_{\text{offset}}=25.263\pm0.075$~lenses horizontal offset and $\text{Y}_{\text{offset}}=0.303\pm0.063$~lenses vertical offset from the FRB 121102 host position (Figure~\ref{fig:kecklris}).\footnote{The FoV offset of 25 pixels in the X direction ($\text{X}_{\text{input}}=25$) is applied using a function of an acquisition unit inside the AO188 instrument. The response to the input signal of $\text{X}_{\text{input}}=25$ pixel on the lenslet array, $\text{X}_{\text{offset}}=25.263\pm0.075$ and $\text{Y}_{\text{offset}}=0.303\pm0.063$, is evaluated using calibration data of a field star obtained during the same Subaru/Kyoto~3DII observation run as the FRB~121102 host galaxy.}
As a flux standard star, G191-B2B was observed with the same instrumental configuration as used for the main target observations.

Data reduction is carried out using custom made IRAF scripts {\tt k3dred} \citep{sug10}\footnote{IRAF is distributed by the National Optical Astronomy Observatory, which is operated by the Association of Universities for Research in Astronomy (AURA) under a cooperative agreement with the National Science Foundation.}.
Briefly, after applying overscan and bias corrections, wavelength-dependent sensitivity variations across the spaxels are corrected using IFU spectra of an internal halogen-lamp of the Kyoto~3DII.
1-D spectra of the target are extracted ``optimally'' \citep{hor86} by utilizing the trace profiles of the internal halogen-lamp spectra as references for extracting traces of the target spectra.
Internal Ne lamp is used for wavelength calibration \citep[see][]{sug10}.
Following \cite{ten17}, Galactic extinction in the direction of FRB~121102 is assumed to be $E(B-V)=0.781$ mag \citep[][]{sch98}, and the observed spectra are corrected for the Galactic extinction using the \cite{car89} Galactic extinction curve.

\section{Results: H$\alpha$ intensity map of the host galaxy of FRB~121102}
\label{sec:halpha_map}

\begin{deluxetable*}{llcccccc}
\tablecaption{Two-dimensional Gaussian fits to the Kyoto~3DII H$\alpha$ map\label{tbl-1}}
\tablewidth{0pt}
\tablehead{
\colhead{} & \colhead{} & \colhead{X} & \colhead{Y} & \colhead{$\sigma_a$} & \colhead{$\sigma_b$} & \colhead{$PA_{\text{XY}}$} & \colhead{Line Flux}\\
\colhead{} & \colhead{} & \colhead{ (pixel) } & \colhead{ (pixel) } & \colhead{ (arcsec) } & \colhead{ (arcsec) } & \colhead{ (degree) } & \colhead{($10^{-16}$ erg/s/cm${}^2$)}
}
\startdata
{$\sigma_a=\sigma_b$}      & FRB 121102 host H$\alpha$ & 19.286$\pm$0.140 & 17.767$\pm$0.170 & 0.238$\pm$0.013 & --- & --- & 2.926$\pm$0.300 \\
{        }                           & Foreground star & 27.953$\pm$0.104 & 17.110$\pm$0.133 & 0.196$\pm$0.010 & --- & --- & ---  \\\hline
{$\sigma_a \neq \sigma_b$} & FRB 121102 host H$\alpha$ & 19.298$\pm$0.136 & 17.755$\pm$0.170 & 0.264$\pm$0.016 & 0.209$\pm$0.015 & 74.01$\pm$12.17 & 2.907$\pm$0.342 \\
{        }                           & Foreground star & 27.966$\pm$0.103 & 17.163$\pm$0.143 & 0.228$\pm$0.019 & 0.176$\pm$0.011 & 82.42$\pm$10.29 & ---  \\
\enddata
\tablecomments{The pixel scale of $0''.0868$/pixel is used when deriving $\sigma_{a}$ and $\sigma_{b}$.}
\end{deluxetable*}

Figure~\ref{fig:each_spaxel} shows the Kyoto~3DII IFU spectra around the spaxel with peak H$\alpha$ intensity $(X,Y)=(19,18)$ at wavelengths around rest-frame H$\alpha$ emission.
Although the H$\alpha$ emission line is clearly detected at the position of the FRB~121102 host galaxy, continuum emission is undetected even after binning over wavelengths because of its faintness.\footnote{Note that binning over wavelengths increases the noise contributions from the readout noise. Therefore, spectroscopic (including IFU) observations are generally less sensitive to continuum emission than broad-band imaging observations.}
The [\ion{S}{2}]$\lambda\lambda 6717, 6731$ doublet is also undetected in our data.
Therefore, below we focus only on the H$\alpha$ emission line.

As indicated in Figure~\ref{fig:each_spaxel}, the H$\alpha$ emission lines are detected at a redshift consistent with $z=0.19273$, and thus our observations independently confirm the redshift determination by \cite{ten17}.
The H$\alpha$ emission line spectrum of each spaxel is fitted at the wavelength range of $\lambda_{\text{obs}}=7650-8000$~\AA\ by a single Gaussian function by fixing the Gaussian width and the observed H$\alpha$ central wavelength as
$\sigma_{\lambda_{\text{obs}}}=3.40$ \AA\ and $\lambda_{\text{obs}}=(1+0.19273)\times 6562.8$~\AA~$=7827.6$~\AA, respectively\footnote{\cite{ten17} showed that the observed-frame line width of the H$\alpha$ emission is $\leq 2.02$~\AA\ at 1$\sigma$. The instrumental spectral resolution of the Subaru/Kyoto~3DII is $2.73$~\AA\ at 1$\sigma$, and therefore we obtain the expected observed line width as $\sigma_{\lambda_{\text{obs}}}\sim\sqrt{2.02^2+2.73^2}=3.40$~\AA, corresponding to the velocity width of $\sigma_v=130$~km/s at $\lambda_{\text{obs}}=7827.6$~\AA. The fitting results are insensitive to the assumption on $\sigma_{\lambda_{\text{obs}}}$.}.
Spectral model fitting is performed using an Interactive Data Language (IDL) routine {\tt MPFIT} \citep{mar09} to minimize the $\chi^2$ values.
The measurement errors of the fitting parameters are estimated by 10,000 trials of Monte Carlo resampling, in which 10,000 mock spectra are generated by adding Gaussian noise to the original spectrum using
the calculated flux density errors \citep[e.g.,][]{and10,she11}.
The H$\alpha$ flux (integrated over the wavelengths) of each spaxel is obtained as a result of fitting.

\begin{figure}[tbp]
\center{
\includegraphics[clip, width=3.2in]{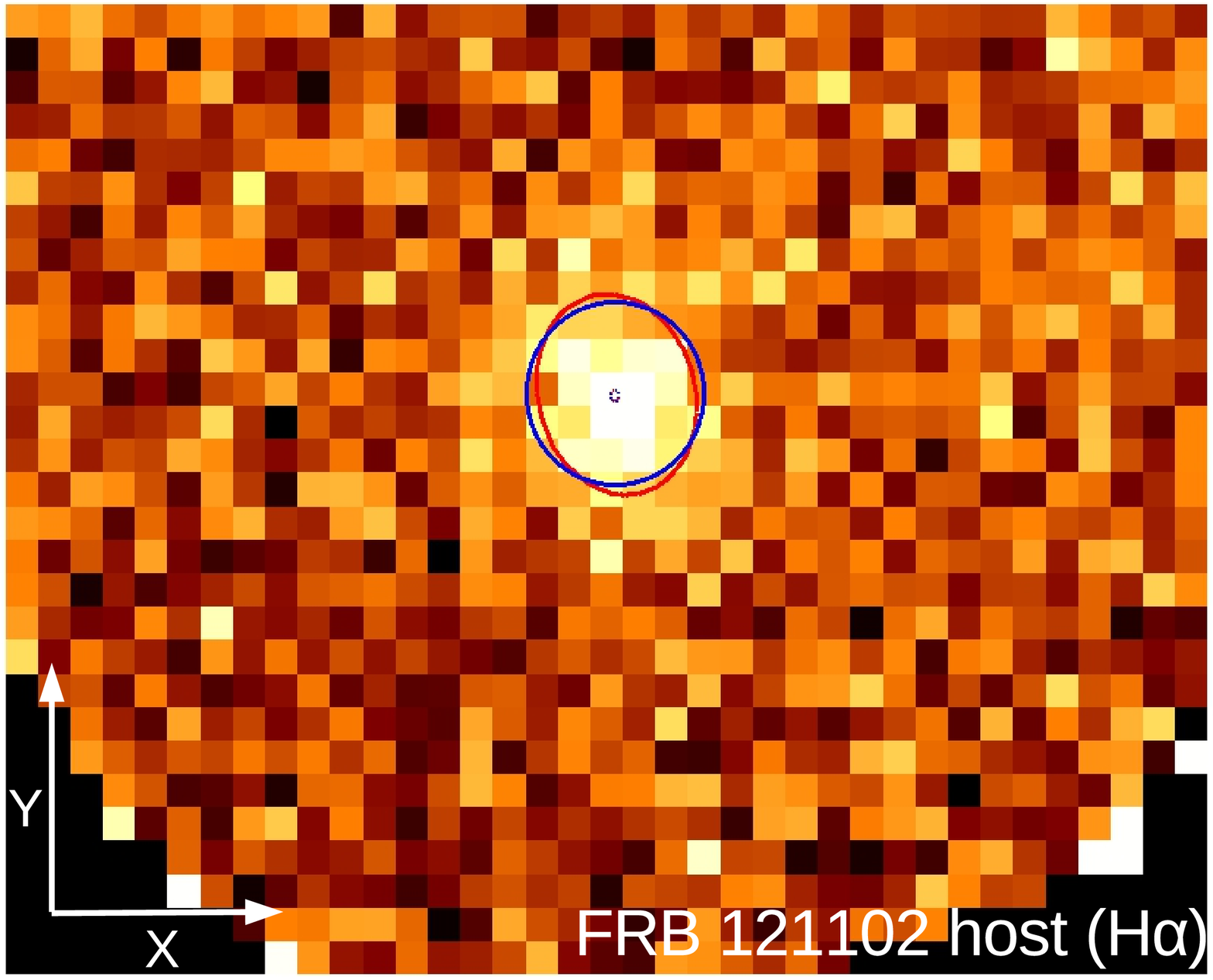}
\includegraphics[clip, width=3.2in]{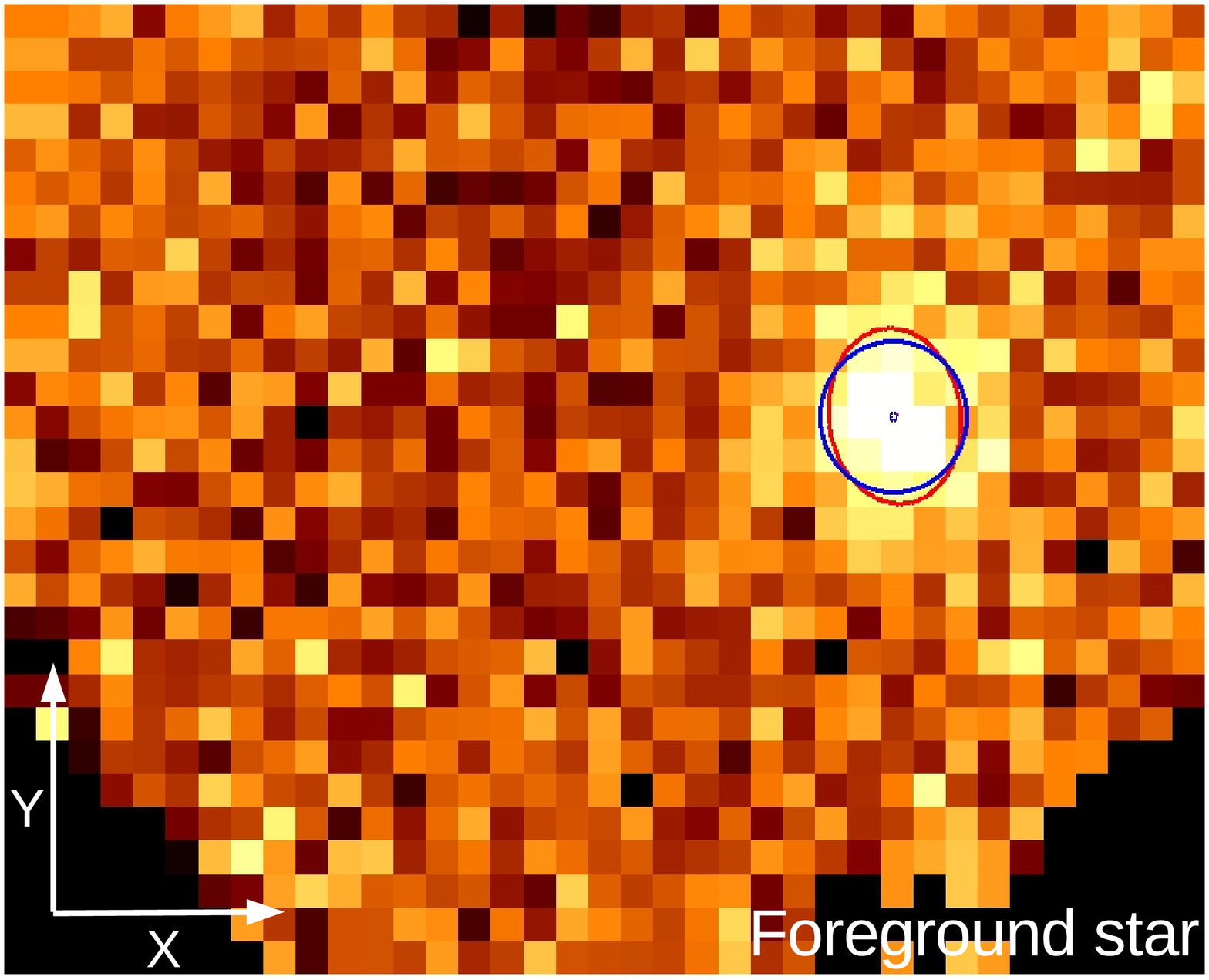}
}
 \caption{
 Kyoto~3DII H$\alpha$ intensity map ($37 \times 29$ pixels = $3''.21 \times 2''.52$) of the host galaxy of FRB~121102 (top) and the no-grism continuum map of the $2.8''$-away foreground star.
 Each pixel corresponds to a spaxel (i.e., the pixel scale is $0''.0868$/pixel).
 Best-fit two-dimensional Gaussians in the cases of $\sigma_a=\sigma_b$ and $\sigma_a \neq \sigma_b$ are also shown as solid circles and ellipses, respectively, where dotted inner ellipses denote the 1$\sigma$ uncertainties of the centroids (see Table~\ref{tbl-1}).
 }
 \label{fig:frb_haplha}
\end{figure}

The upper panel of Figure~\ref{fig:frb_haplha} shows the spatial profile (intensity map) of the H$\alpha$ flux of the host galaxy of FRB~121102.
It should be noted that the same single Gaussian model is fitted to all $37 \times 29$ spaxels, and thus the best-fit H$\alpha$ flux has non-zero values even at spaxels outside the H$\alpha$ emission region due to the noise.
As can be clearly seen in Figure~\ref{fig:frb_haplha}, our Subaru/Kyoto~3DII IFU observations reveal that there is only a single compact H$\alpha$ emission region in the host galaxy of FRB~121102.

Then, two-dimensional symmetric ($\sigma_a=\sigma_b$) and asymmetric ($\sigma_a \neq \sigma_b$) Gaussians are fitted to the H$\alpha$ spatial profile shown in Figure~\ref{fig:frb_haplha}, where $\sigma_a$ and $\sigma_b$ denote the Gaussian widths in the directions of semi-major ($a$) and semi-minor ($b$) axes, respectively.
The best-fit two-dimensional Gaussians to the H$\alpha$ spatial profile are shown in Figure~\ref{fig:frb_haplha}, and Table~\ref{tbl-1} lists the best-fit parameters, where $X$ and $Y$ denote the centroid positions, ``$PA_{\text{XY}}$'' denotes the position angle of the asymmetric ($\sigma_a \neq \sigma_b$) Gaussian measured clockwise relative to the $X$-axis, and ``Line Flux'' denotes the H$\alpha$ line flux integrated over the Gaussian profiles.
The same two-dimensional Gaussians are also fitted to the no-grism continuum image of the foreground star (see Table~\ref{tbl-1}).

The measured Gaussian widths of the spatial profile of the H$\alpha$ emission region are slightly larger than the foreground star (i.e., point source).
If we take the fitting results in the case of the $\sigma_a = \sigma_b$ model (Table~\ref{tbl-1}) at face value, the size of the H$\alpha$ emission region can be calculated as $\sqrt{0''.238^2-0''.196^2}\sim0''.14$ at 1$\sigma$ radius or $0''.33$ at FWHM.
However, taking the time-variability of the natural seeing conditions between exposures and the stability of the AO-correction into account, this slight difference does not necessarily indicate that our observations detected an extended structure of the H$\alpha$ emission region.
Although the asymmetric Gaussian fit for the H$\alpha$ emission region results in a slightly elongated spatial profile, it should be noted that the same level of elongation of the spatial profile is also observed in the case of the no-grism continuum map of the foreground star.
Therefore, we conservatively consider that the H$\alpha$ emission region in the FRB~121102 host galaxy is not spatially-resolved or only marginally resolved by our AO-assisted Kyoto~3DII IFU observation, indicating that the observed profile of the H$\alpha$ emission region in Figure~\ref{fig:frb_haplha} is mostly the point spread function (PSF) of the image.
Below, we adopt the best-fit parameters from the two-dimensional symmetric Gaussian fit ($\sigma_a = \sigma_b$) tabulated in Table~\ref{tbl-1} as the spatial profile parameters of the H$\alpha$ emission region in the FRB~121102 host galaxy.
From the values of $\sigma_{a}$ shown in Table~\ref{tbl-1}, we obtained a stringent upper limit for the angular size of the H$\alpha$ emission region as 
$r_{H\alpha} < 0''.24$ (0.79~kpc) at 1$\sigma$ radius or $<$~$0''.57$ (1.9~kpc) at FWHM.
It should be noted that, as mentioned above, the size parameter of the H$\alpha$ emission region may be as small as $r_{H\alpha} \sim 0''.14$ (0.46~kpc) at 1$\sigma$ radius.

The measured ``Line Flux'' integrated over the two-dimensional symmetric Gaussian profile is $2.926~(\pm 0.300) \times 10^{-16}$ erg/s/cm${}^2$ (Table~\ref{tbl-1}), which is consistent with the spatially-unresolved H$\alpha$ line flux measurement reported by \cite{ten17} from Gemini/GMOS $1''.0$-width long-slit spectrum \citep[see Table~1 of][]{ten17}.
This consistency indicates that the H$\alpha$ emission from the host galaxy of FRB~121102 is confined only to the compact H$\alpha$-emitting clump revealed by our observations, and there is no significant H$\alpha$ flux loss in the Gemini/GMOS long-slit spectrum.
As the measurement of the H$\alpha$ emission line flux by \cite{ten17} is better calibrated and has a higher signal-to-noise ratio than our measurement, we adopt the value of $F_{\text{H}\alpha}=2.608~(\pm0.036)\times10^{-16}~\text{erg/s/cm}^2$ (Galactic extinction is corrected) reported by \cite{ten17} as the integrated H$\alpha$ line flux.
It should be noted that the Galactic extinction-corrected Balmer decrement H$\alpha$/H$\beta=2.714 \pm 0.256$ measured from the Gemini/GMOS spectrum is consistent with theoretical values for Case~B recombination in photoionized nebulae \citep[Table~4.4 of][]{ost06}, suggesting that the dust extinction in the host galaxy of FRB~121102 is negligible.

\section{Discussion}
\label{discuss}

\subsection{Positional offsets between the centroids of the H$\alpha$ emission region, stellar continuum, and FRB~121102 radio bursts}
\label{sky_coordinate}

The relative pixel coordinate from the continuum emission of the $2''.8$-away foreground star to the H$\alpha$ emission region in the FRB~121102 host is calculated from the best-fit parameters of the two-dimensional symmetric ($\sigma_a=\sigma_b$) Gaussian fits tabulated in Table~\ref{tbl-1} as
\begin{eqnarray}
&&(\Delta\text{X}\ [\text{pixel}], \Delta\text{Y}\ [\text{pixel}])\nonumber\\
&\equiv& (\text{X}_{\text{FRB\ host\ H}\alpha} -\text{X}_{\text{Foreground star}} -\text{X}_{\text{offset}},  \nonumber\\ 
& &\text{Y}_{\text{FRB\ host\ H}\alpha} - \text{Y}_{\text{Foreground star}}-\text{Y}_{\text{offset}})\nonumber\\
&=& (-33.930 \pm 0.190, 0.354 \pm 0.225) \label{rel_pixel}
\end{eqnarray}
where the factors $\text{X}_{\text{offset}}$ and $\text{Y}_{\text{offset}}$ represent the relative FoV offset between exposures of the FRB~121102 host and the foreground star (see Section~\ref{observation}).
The relative sky coordinates between the positions of the H$\alpha$ emission region in the FRB~121102 host and the foreground star (i.e., $\Delta \text{R.A.}$ and $\Delta \text{Dec.}$, respectively), and their 1$\sigma$ uncertainties, can be calculated from Equation~\ref{rel_pixel}, the on-sky $PA$ of $108.99\pm0.16$~deg, and the pixel scale of $0''.0868\pm0''.0002$~/pixel (see Figure~\ref{fig:kecklris}) as:
\begin{eqnarray}
&&(\Delta \text{R.A.}, \Delta \text{Dec.}) \nonumber \\
&=&  (0''.9874 \pm 0''.0209, 2''.7748 \pm 0''.0182)\ \ \ [\text{H}\alpha].
\label{relative_ha}
\end{eqnarray}

\cite{ten17} derived the sky coordinates (J2000.0) of the foreground star measured on the Gemini/GMOS $z'$-band images as
\begin{eqnarray}
& &(\text{R.A.}, \text{Dec.}) \nonumber\\
&=&(05^{\text{h}}31^{\text{m}}58^{\text{s}}.6180 \pm 8.6~\text{mas}, +33^{\circ}08^{'}49^{''}.837 \pm 9.3~\text{mas}), \nonumber
\end{eqnarray}
and of the FRB~121102 host galaxy measured on the Gemini/GMOS $r'$-, $i'$-, and $z'$-band images as
\begin{eqnarray}
& &(\text{R.A.}, \text{Dec.}) \nonumber\\
&=&(05^{\text{h}}31^{\text{m}}58^{\text{s}}.6876 \pm 50.8~\text{mas}, +33^{\circ}08'52''.490 \pm 46.4~\text{mas}) \nonumber\\
&=&(05^{\text{h}}31^{\text{m}}58^{\text{s}}.6895 \pm 30.6~\text{mas}, +33^{\circ}08'52''.466 \pm 25.8~\text{mas}) \nonumber\\ 
&=&(05^{\text{h}}31^{\text{m}}58^{\text{s}}.6833 \pm 43.5~\text{mas}, +33^{\circ}08'52''.379 \pm 43.5~\text{mas}), \nonumber
\end{eqnarray}
respectively (S. Tendulkar \& C. Bassa, personal communications).
The coordinates of the FRB~121102 host galaxy are the positions of two-dimensional asymmetric Gaussians fitted to the GMOS $r'$, $i'$, and $z'$-band images.
As the FRB~121102 host galaxy may have an irregular shape intrinsically, these positions are considered to be the centroid positions of the broad-band light \citep[][]{ten17}.
\cite{mar17} derived the average position of four FRB~121102 radio bursts detected by the European Very Long Baseline Interferometry Network on September 20, 2016 (see their Table~1), as
\begin{eqnarray}
& &(\text{R.A.}, \text{Dec.}) \nonumber\\
&=&(05^{\text{h}}31^{\text{m}}58^{\text{s}}.70119 \pm 4~\text{mas},+33^{\circ}08'52''.5536 \pm 2.2~\text{mas}).
\nonumber
\end{eqnarray}
It should be noted that a variable persistent non-thermal radio source with a size of $<$ 0.2 mas (0.7~pc) has also been identified at the same position as the FRB~121102 radio bursts \citep[][]{cha17,mar17}, which is believed to have a direct relationship to the FRB~121102 radio bursts \citep[see e.g.,][]{mur16,met17,kas17,bel17}.
From these values, the relative sky coordinates of the positions of the $r'$-, $i'$-, and $z'$-band light of the FRB~121102 host and the FRB~121102 radio bursts measured from the foreground star are calculated as
\begin{eqnarray}
&&(\Delta \text{R.A.}, \Delta \text{Dec.})\nonumber \\
&=&(0''.8741 \pm 0''.0515, 2''.6530 \pm 0''.0473)\ \ \ [r'\text{-band}]\label{relative_cont_r}\\
&=&(0''.8980 \pm 0''.0318, 2''.6290 \pm 0''.0274)\ \ \ [i'\text{-band}]\label{relative_cont_i}\\
&=&(0''.8201 \pm 0''.0443, 2''.5420 \pm 0''.0445)\ \ \ [z'\text{-band}]\label{relative_cont_z}\\
&=&(1''.0448 \pm 0''.0095, 2''.7166 \pm 0''.0096)\ \ \ [\text{Radio Burst}].
\label{relative_cont_radio}
\end{eqnarray}

\begin{figure}[tbp]
\center{
\includegraphics[clip, width=3.2in]{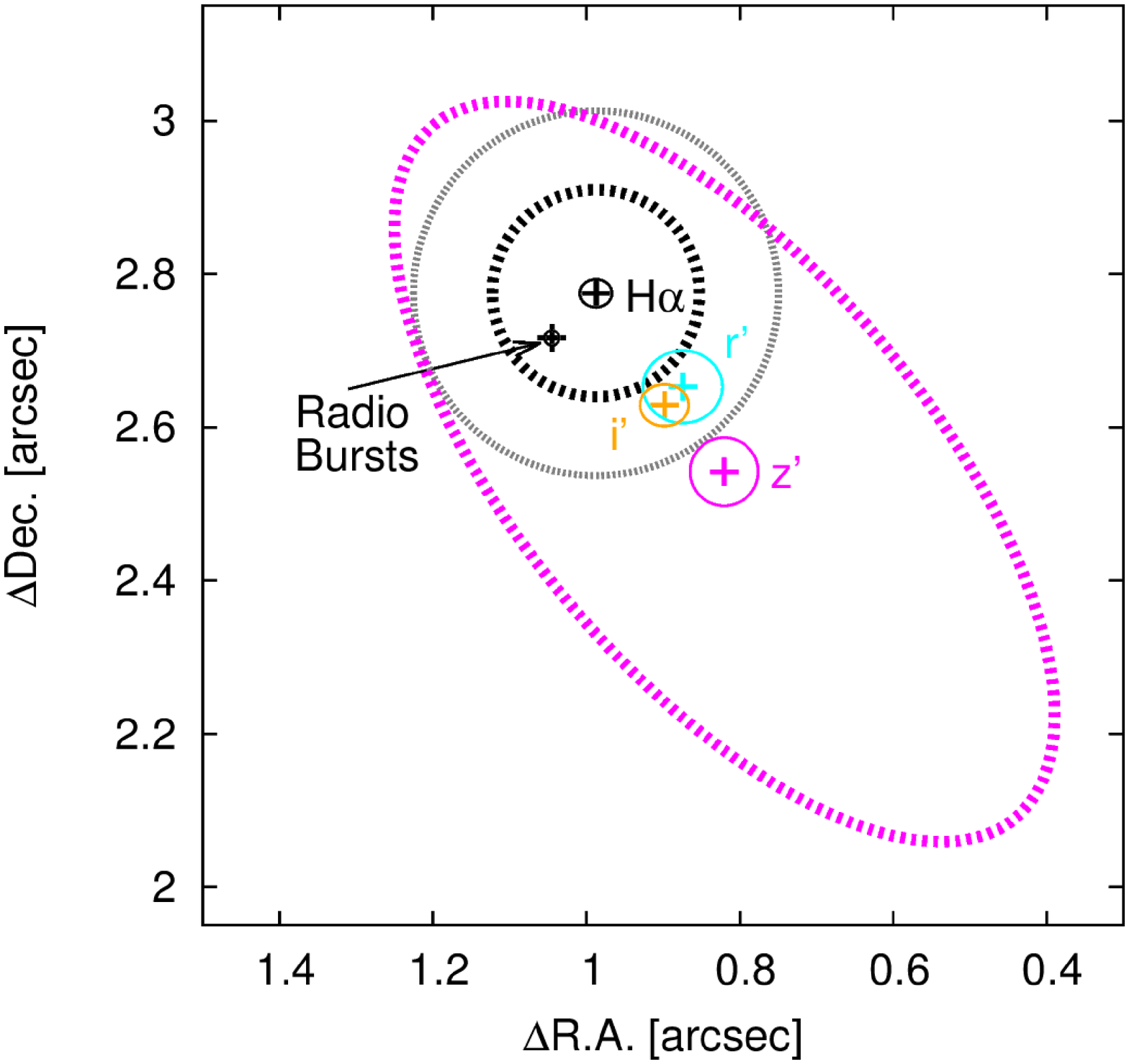}
}
 \caption{
 The relative sky coordinates (measured from the foreground star) of the centroid positions of the H$\alpha$ emission (this work), and of the $r'$-band, $i'$-band, and $z'$-band light \citep[Gemini/GMOS;][]{ten17} of the FRB~121102 host (crosses).
 Small thin ellipses denote the 1$\sigma$ absolute positional uncertainties.
 The larger thick dotted ellipse denotes the two-dimensional Gaussian fit for the stellar continuum-dominated $z'$-band image \citep[Figure~3 of][]{ten17}.
The thick and thin dotted circles around the centroid position of the H$\alpha$ emission indicate the estimated size of the H$\alpha$ emission region in the FRB~121101 host galaxy ($r_{\text{H}\alpha} \sim 0''.14$ and a more conservative value, $r_{\text{H}\alpha} < 0''.24$, respectively, at 1$\sigma$ radius; see Section~\ref{sec:halpha_map}).
 }
 \label{fig:continuum_halpha_coordinates}
\end{figure}

The relationships among the centroid positions of the H$\alpha$ emission region, $r'$-, $i'$-, and $z'$-band light, and the FRB~121102 bursts are shown in Figure~\ref{fig:continuum_halpha_coordinates}.
As noted by \cite{ten17}, the $z'$-band light is dominated by the stellar continuum light, and $r'$ and $i'$-bands contain both the stellar continuum light and the strong line emission (H$\beta$ and [\ion{O}{3}] in $r'$-band, and H$\alpha$ in $i'$-band).
Therefore, based on the observed offsets between the centroids of the $r'$-, $i'$-, and $z'$-band light, \cite{ten17} suggested that the FRB~121102 host galaxy has an \ion{H}{2} region at a slight offset from the centroid of the stellar continuum.
Our observations verify this suggestion; the centroid of the H$\alpha$ emission region determined from our Kyoto~3DII data is located at a position offset from the centroid of the $z'$-band light by $0''.29 \pm 0''.05$ ($ 0.96 \pm 0.17$~kpc).
On the other hand, the spatial offset between the centroid of the H$\alpha$ emission region and the radio burst position, denoted as $d_{\text{FRB}-\text{H}\alpha}$, is quite small [$d_{\text{FRB}-\text{H}\alpha}=0''.08 \pm 0''.02$ ($0.26\pm0.07$~kpc)].
As the the spatial extent of the H$\alpha$ emission region is $r_{\text{H}\alpha} \sim 0''.14$ (or more conservatively, $r_{\text{H}\alpha} < 0''.24$; Section~\ref{sec:halpha_map}), we conclude that the FRB~121102 radio bursts are located within the H$\alpha$ emission region.

Based on observations, hydrogen-poor SLSNe and LGRBs are known to preferentially occur in the brightest star-forming regions in their (low-metallicity) host galaxies, where the most massive stars are expected to form \citep[][and references therein]{fru06,kel08,lun14,ang16,bla16,nii17,lym17}.
As described above, FRB~121102 shows a similar spatial association with the H$\alpha$ emission (i.e., star-forming) region in the host galaxy, suggesting that FRB~121102 radio bursts are produced by a young neutron star or a magnetar formed as a result of a massive star explosion, such as a hydrogen-poor SLSN or LGRB \citep[e.g.,][]{met17}.
Therefore, our observations provide further observational evidence for young pulsar or magnetar models for FRB~121102 \citep[e.g.,][]{kul15,mur16,lyu16b,met17,nic17}.

It should also be noted that the observed H$\alpha$ emission region may point to the location of an active galactic nucleus (AGN) of the host galaxy of FRB~121102, which is offset from the centroid of the stellar light.
BPT diagnostics \citep{bal81} for the emission lines of the FRB~121102 host galaxy investigated by \cite{ten17} are consistent with star-formation origin, but as it is observationally known that many radio-loud AGNs show only weak or no optical emission line signatures of AGN activity \citep[][and references therein]{mau07,ten17}, we cannot conclusively rule out the possibility that the H$\alpha$ emission region is (partly) photoionized by the AGN of the FRB~121101 host galaxy.
If so, the slight (but statistically significant) spatial offset between the centroid of the H$\alpha$ emission region and the FRB~121102 radio bursts described above ($d_{\text{FRB}-\text{H}\alpha} \sim 0''.08 = 0.26$~kpc) may suggest that FRB~121102 and the associated persistent radio source are not directly related to the AGN activities.

\subsection{The DM contribution of the H$\alpha$ emission region to the observed DM}
\label{dm_contribution}

Based on the observations of multiple radio bursts from FRB~121102,  \cite{spi16} reported the $1\sigma$ range of the observed DMs of the repeating radio bursts of FRB~121102 as $\text{DM}_{\text{obs}}=558.1 \pm 3.3~\text{pc}~\text{cm}^{-3}$.
As DM is an additive quantity, the DM contributions from different regions to the observed DM (denoted as $\text{DM}_{\text{obs}}$) can be divided into each component as
\begin{equation}
\text{DM}_{\text{obs}} = \text{DM}_{\text{MW}} + \text{DM}_{\text{IGM}} + \text{DM}_{\text{host, obs}}, 
\label{dm_add}
\end{equation}
where $\text{DM}_{\text{MW}}$, $\text{DM}_{\text{IGM}}$, and $\text{DM}_{\text{host, obs}}$ represent the DM contributions from the Milky Way (MW), ionized baryons in IGM, and the host galaxy of FRB~121102, respectively \citep[e.g.,][]{xu15,ten17,yan17}.

According to the NE2001 model \citep{cor02}, the MW contribution is $\text{DM}_{\text{MW}}=218\ \text{pc}\ \text{cm}^{-3}$ in the direction of FRB~121102 with uncertain errors of the order of 20\% \citep[e.g.,][]{abd13,ten17}.
The DM contribution from the IGM to the observed DM of FRB~121102 at $z=0.19273$ can be calculated as \citep[][]{den14,kea16,yan17}:
\begin{eqnarray}
\text{DM}_{\text{IGM}} &\simeq& 177~\text{pc}~\text{cm}^{-3} \times  \left(\frac{\Omega_{\text{IGM}}}{0.044}\right) \left(\frac{H_0}{67.7}\right)\label{dm_igm_eq1}\\
\Omega_{\text{IGM}} &\equiv& 0.044 \left(\frac{f_{\text{IGM}}}{0.9}\right) \left(\frac{\Omega_{\text{b}}}{0.049}\right)\label{dm_igm_eq2}
\end{eqnarray}
where $\Omega_{\text{IGM}}$ and $f_{\text{IGM}}$ are the cosmic density of ionized baryons in the IGM and the fraction of baryon mass in the IGM \citep[][]{fuk04,shu12}\footnote{We assume that hydrogen and helium are fully ionized at $z<2$ \citep[][and references therein]{den14}.}, respectively.
The associated standard deviation of $\text{DM}_{\text{IGM}}$ is $\approx 85~\text{cm}^{-3}$ \citep{mcq14,ten17}.
Therefore, the DM contribution from the FRB host is in the range of
\begin{equation}
\text{DM}_{\text{host, obs}} = 163 \pm 96~\text{pc}\ \text{cm}^{-3},
\label{dm_host}
\end{equation}
where the uncertainty of $\pm 96~\text{pc}\ \text{cm}^{-3}$ is evaluated by adding the uncertainties of $\text{DM}_{\text{MW}}$, $\text{DM}_{\text{IGM}}$, and $\text{DM}_{\text{obs}}$ in quadrature.

\begin{figure}[tbp]
\center{
\includegraphics[clip, width=3.4in]{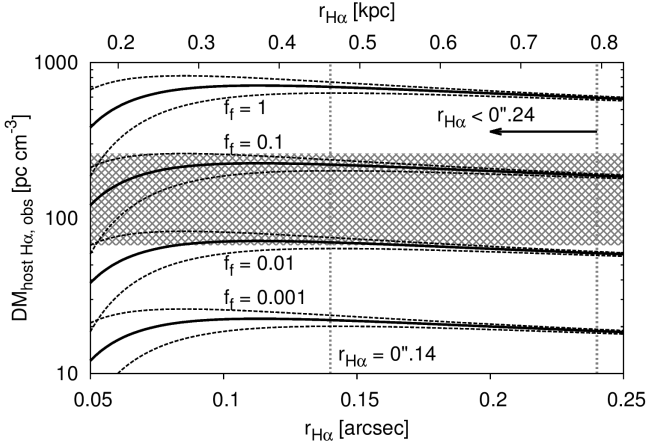}
}
 \caption{
 Observational constraints on the DM contribution from the observed H$\alpha$ emission region (Equation~\ref{eqn_dm3}) for $f_{\text{f}}=1$, $0.1$, $0.01$, and $0.001$, where $C$ is fixed to 1. The thick solid lines are for $d_{\text{FRB}-\text{H}\alpha}=0''.08$, and the upper and lower dashed lines associated with each of the thick solid lines are for $d_{\text{FRB}-\text{H}\alpha}=0''.06$ and $d_{\text{FRB}-\text{H}\alpha}=0''.10$, respectively. The hatched region denotes $\text{DM}_{\text{host, obs}}=163 \pm 96$~pc~cm${}^{-3}$ (Equation~\ref{dm_host}).
 The vertical lines denote the estimated size of the H$\alpha$ emission region (see Figure~\ref{fig:continuum_halpha_coordinates}).
 }
 \label{fig:dm_figure}
\end{figure}

It is worth examining whether the compact H$\alpha$ emission region in the host galaxy of FRB~121102 detected by our observations can have a significant DM contribution on $\text{DM}_{\text{host, obs}}$ of FRB~121102 \citep[see e.g.,][]{xu15,kul15}.
The surface brightness of the H$\alpha$ emission line provides an estimate of the DM contribution from H$\alpha$-emitting gas in the host galaxy of FRB~121102 \citep[][and references therein]{kul14,kul15,sch16,ten17,yan17}.
Here, we use the stringent upper limit of $r_{\text{H}\alpha} = 0''.24$ (Section~\ref{sec:halpha_map}) as a reference value for the size of the H$\alpha$ emission region, but it should be noted that the final results are insensitive to the exact value of $r_{\text{H}\alpha}$ (see Figure~\ref{fig:dm_figure} and Equation~\ref{eqn_dm4}).
From the Galactic extinction-corrected line flux $F_{\text{H}\alpha}=2.608 \times 10^{-16} \text{erg/s/cm}^2$ (Section~\ref{sec:halpha_map}), H$\alpha$ surface brightness $S(\text{H}\alpha)_\text{host, obs}$ is
\begin{eqnarray}
S(\text{H}\alpha)_\text{host, obs} &=& \frac{F_{\text{H}\alpha}}{\pi r_{\text{H}\alpha}^2}\nonumber\\
&\approx& 1.441 \times 10^{-15} \ \ \ \ \text{erg}\ \text{cm}^{-2}\ \text{s}^{-1}\ \text{arcsec}^{-2} \times \left( \frac{r_{\text{H}\alpha}}{0''.24} \right)^{-2}\nonumber
\end{eqnarray}
H$\alpha$ surface brightness in the source frame $S(\text{H}\alpha)_\text{host, s}$ is
\begin{eqnarray}
S(\text{H}\alpha)_\text{host, s} &=& (1+z)^4 S(\text{H}\alpha)_\text{host, obs}\nonumber\\
&\approx& 2.916 \times 10^{-15} \ \ \ \ \text{erg}\ \text{cm}^{-2}\ \text{s}^{-1}\ \text{arcsec}^{-2} \times \left( \frac{r_{\text{H}\alpha}}{0''.24} \right)^{-2}.\nonumber
\end{eqnarray}
For Case~B recombination in hydrogen plasma with temperature $T=10000$~K, $S(\text{H}\alpha)_\text{host, s}$ is related to source frame emission measure (denoted as $\text{EM}_\text{host, s}$) as \citep{rey77,ten17}
\begin{eqnarray}
\text{EM}_{\text{host, s}} &=& 486~\text{pc}~\text{cm}^{-6} \left( \frac{T}{10000~\text{K}} \right)^{0.9}\nonumber\\
& &\times \left( \frac{S(\text{H}\alpha)_\text{host, s}}{10^{-15}~\text{erg}\ \text{cm}^{-2}\ \text{s}^{-1}\ \text{arcsec}^{-2}} \right)\nonumber\\
&\approx& 1417~\text{pc}~\text{cm}^{-6} \times \left( \frac{r_{\text{H}\alpha}}{0''.24} \right)^{-2}\label{emission_measure}
\end{eqnarray}

By considering an ionized gas clump (with the volume-averaged electron density $n_{e_{c}}$), which consists of sub-clumps with internal density $n_{e_{sc}}$ and filling factor $f_{\text{f}} \leq 1$ ($f_{\text{f}} \equiv n_{e_{c}}/ n_{e_{sc}}$), EM and DM are related to each other as follows \citep{cor16b,ten17}
\begin{eqnarray}
\text{DM}_{\text{host H}\alpha\text{, s}} &=& 387\ \text{pc}\ \text{cm}^{-3} \times f_{\text{f}}^{1/2}C \left(\frac{L}{1\ \text{kpc}}\right)^{1/2} \left( \frac{\text{EM}_{\text{host, s}}}{600\ \text{pc}\ \text{cm}^{-6}} \right)^{1/2}\label{eqn_dm}\\
 C &=& \left(\frac{1}{\zeta (1+\epsilon^2)/4}\right)^{1/2} \leq 2,
\end{eqnarray}
where $\epsilon\equiv$(RMS density) / (mean density) $\geq 0$ represents small scale fractional density fluctuation within each sub-clump (RMS = root mean square),  $\zeta \geq 1$ is the dimensionless second moment of the density variations between sub-clumps, and $L$ is the path length through the clump \citep[see][for details]{cor16b}.
$\text{DM}_{\text{host H}\alpha\text{, s}}$ denotes the DM contribution from the H$\alpha$ emission region in the FRB~121102 host galaxy.
As assumed by \cite{ten17}, we adopt $\zeta=2$ for the density variations between sub-clumps and $\epsilon=1$ for the density fluctuation within each sub-clump (i.e., $C=1$) as a reference value.
The observed-frame DM contribution from the H$\alpha$ emission region $\text{DM}_{\text{host H}\alpha\text{, obs}}$ can be evaluated by multiplying dimensionless factors of $1/(1 + z)$ and $L_{\text{FRB}}/L$ to $\text{DM}_{\text{host H}\alpha\text{, s}}$ \citep{ten17}, where $L_{\text{FRB}}$ indicates an effective path length through the clump:
\begin{eqnarray}
\text{DM}_{\text{host H}\alpha\text{, obs}} &=& \left( \frac{1}{1+z} \right) \left(\frac{L_{\text{FRB}}}{L}\right) \text{DM}_{\text{host H}\alpha\text{, s}} \nonumber\\
 &=& 324\ \text{pc}\ \text{cm}^{-3} \times \left(\frac{L_{\text{FRB}}}{L}\right) f_{\text{f}}^{1/2}C \left(\frac{L}{1\ \text{kpc}}\right)^{1/2}  \nonumber\\
&\times&\left( \frac{\text{EM}_{\text{host, s}}}{600\ \text{pc}\ \text{cm}^{-6}} \right)^{1/2}
\label{eqn_dm2}
\end{eqnarray}

Our Kyoto~3DII IFU observations revealed that the H$\alpha$ emission region with the Gaussian width of $r_{\text{H}\alpha} < 0''.24$ is located at a position offset from the FRB~121102 position by $d_{\text{FRB}-\text{H}\alpha}=0''.08 \pm 0''.02$ (Sections~\ref{sec:halpha_map} and \ref{sky_coordinate}).
This result places observational constraints on $L$ and $L_{\text{FRB}}/L$ in Equation~\ref{eqn_dm2} as follows.
$L$ can be evaluated as $L=2r_{\text{H}\alpha}$ with the assumption that the H$\alpha$ emission region is spherical, and $L_{\text{FRB}}/L$ can be evaluated as $G(r=d_{\text{FRB}-\text{H}\alpha})/G(r=0''.00)$ [where $G(r)$ is the Gaussian profile centered on the H$\alpha$ emission region with the Gaussian width of $r$].
By substituting these values and Equation~\ref{emission_measure} into Equation~\ref{eqn_dm2}, $\text{DM}_{\text{host H}\alpha\text{, obs}}$ (normalized at $d_{\text{FRB}-\text{H}\alpha}=0''.08$ and $r_{\text{H}\alpha}=0''.24$) can be expressed as a function of $r_{\text{H}\alpha}$
\begin{eqnarray}
\text{DM}_{\text{host H}\alpha\text{, obs}} = 594\ \text{pc}\ \text{cm}^{-3} \times f_{\text{f}}^{1/2}C \times \left( \frac{r_{\text{H}\alpha}}{0''.24} \right)^{-1/2} \nonumber\\
 \times  \left[ \exp \left(-\frac{(d_{\text{FRB}-\text{H}\alpha})^2}{2(r_{\text{H}\alpha})^2}\right) \bigg/ \exp \left(-\frac{(0''.08)^2}{2(0''.24)^2}\right) \right].
\label{eqn_dm3}
\end{eqnarray}

Figure~\ref{fig:dm_figure} shows Equation~\ref{eqn_dm3} as a function of $r_{\text{H}\alpha}$.
As can be seen in this figure, the DM contribution from the FRB host galaxy $\text{DM}_{\text{host, obs}}=163 \pm 96$~pc~cm${}^{-3}$ (Equation~\ref{dm_host}) can fully be attributed to the observed H$\alpha$ emission region under the observational constraint of $r_{\text{H}\alpha}<0''.24$ (or $r_{\text{H}\alpha}\sim0''.14$; see Section~\ref{sec:halpha_map}) if the filling factor takes large values of $f_{\text{f}}\simeq0.01-0.1$ and the linear size of the H$\alpha$ emission region is larger than $r_{\text{H}\alpha}~\gtrsim~0''.1$ ($0.3$~kpc).
These values of $f_{\text{f}}$ and $r_{\text{H}\alpha}$ are within the range expected for low-density \ion{H}{2} regions \citep{pyn93,pyn16,ced13}.
Thus, our observations indicate that $\text{DM}_{\text{host H}\alpha\text{, obs}}$ can make a major contribution to $\text{DM}_{\text{host, obs}}$, and DM contributions from other DM sources, such as a young supernova remnant around the progenitor of FRB~121102 \citep[e.g.,][]{pir16,yan16,yan17,kas17}, are probably small.
The absence of the significant DM contribution from the supernova remnant is consistent with the time-constancy of the observed DM of FRB~121102 (Section~\ref{sec:intro}), which requires that the supernova remnant, if present, should be older than $\sim$ 100 years and thus have little DM contribution of $\ll$~100~pc~cm${}^{-3}$ \citep[][]{mur16,pir16,ten17,cao17,met17}.

Alternatively, we can use Equation~\ref{eqn_dm3} to derive an observational constraint on $\Omega_{\text{IGM}}$.
Equation~\ref{eqn_dm3} as a function of $r_{\text{H}\alpha}$ takes a maximum value at $r_{\text{H}\alpha} = \sqrt{2} d_{\text{FRB}-\text{H}\alpha}$, and thus the maximum possible DM contribution from the H$\alpha$ emission region (regardless of the exact value of $r_{\text{H}\alpha}$) can be derived algebraically as:
\begin{eqnarray}
&&\text{DM}_{\text{host H}\alpha\text{, obs}} (r_{\text{H}\alpha} = \sqrt{2} d_{\text{FRB}-\text{H}\alpha}) \nonumber\\
&=& 712~\text{pc~cm}{}^{-3} \times f_{\text{f}}^{1/2} C \times \left(\frac{d_{\text{FRB}-\text{H}\alpha}}{0''.08}\right)^{-1/2}\nonumber\\
&=& 712~(\pm~89)~\text{pc~cm}{}^{-3} \times f_{\text{f}}^{1/2} C,
\label{eqn_dm4}
\end{eqnarray}
where the uncertainty of $\pm 89~\text{pc~cm}{}^{-3}$ is from the measurement error in $d_{\text{FRB}-\text{H}\alpha}$.
If we assume that $f_{\text{f}}$ and $C$ should be less than $0.1$ and $1$, respectively, under real physical conditions, and that $\text{DM}_{\text{host, obs}}$ is solely due to the DM contribution from the observed H$\alpha$ emission region, $\text{DM}_{\text{host, obs}}$  can only take a value less than $225~\pm~28$~pc~cm${}^{-3}$.
Thus, the DM contribution from the IGM ($\text{DM}_{\text{IGM}}$) can be constrained from Equation~\ref{dm_add} as $\text{DM}_{\text{IGM}} >48~\text{pc}~\text{cm}^{-3}$ at the 90\% confidence lower limit.
By substituting this constraint into Equation~\ref{dm_igm_eq1}, we obtain a 90\% confidence level lower limit on the cosmic baryon density in the IGM in the low-redshift universe as
\begin{equation}
\Omega_{\text{IGM}}~>~0.012,
\label{dm_igm}
\end{equation}
or equivalently, $f_{\text{IGM}} > 0.25$ (Equation~\ref{dm_igm_eq2}).
According to the estimate of the cosmic baryon budget in the low redshift universe within the $\Lambda$CDM cosmology \citep[][]{fuk04b,fuk04,shu12}, approximately 90~\% of the baryons must be lying in the IGM (as assumed in Equation~\ref{dm_igm_eq2}), which have been historically called `missing' baryons \citep[for a review, see][]{bre07}.
The observational constraints on the geometry of the H$\alpha$ emission region in the FRB~121102 host galaxy obtained by our observations suggest that ionized baryons in the IGM make up at least 25~\% of the baryonic mass of the universe at $z \sim 0$, and thus provide evidence for the current best estimate of the cosmic baryon budget.

\section{Conclusions}
\label{conclusion}

Here, we present the H$\alpha$ intensity map of the host galaxy of the repeating fast radio burst FRB~121102 obtained with AO-assisted Kyoto~3DII IFU.
Our observations independently confirm the results of \cite{ten17} indicating that the host galaxy of FRB~121102 is a star-forming dwarf galaxy at a redshift of $z\simeq0.193$ (Figure~\ref{fig:each_spaxel}).
The detected H$\alpha$ emission region in the host galaxy is not spatially-resolved or only marginally resolved, and therefore we obtain an upper size limit $r_{\text{H}\alpha}\lesssim 0''.24$ (1$\sigma$ radius), and there is an indication that the size of the H$\alpha$ emission region is as small as $0''.14$ (1$\sigma$ radius).
The H$\alpha$ emission region is located at a position offset from the extended ($\simeq 1''.0-1''.4$) stellar continuum emission region by $0''.29 \pm 0''.05$ (Figure~\ref{fig:continuum_halpha_coordinates}).
The spatial offset between the centroid of the H$\alpha$ emission region and the FRB~121102 radio bursts is $d_{\text{FRB}-\text{H}\alpha}=0''.08 \pm 0''.02$, indicating that FRB~121102 reside in the observed H$\alpha$ emission region (Figure~\ref{fig:continuum_halpha_coordinates}).
The spatial association between the H$\alpha$-emitting star-forming region in the host galaxy and the position of FRB~121102 suggests that FRB~121102 is produced by a young pulsar/magnetar formed by a massive star explosion.

Our observations confirm that the sightline toward FRB~121102 passes through the H$\alpha$-emitting clump, and thus the H$\alpha$ emission region can make a major contribution to the host galaxy DM of FRB~121102 ($\text{DM}_{\text{host, obs}}=163 \pm 96$~pc~cm${}^{-3}$) under reasonable assumptions regarding the physical properties of the H$\alpha$ emission region (Equation~\ref{eqn_dm3} and Figure~\ref{fig:dm_figure}).
Conversely, observational constraints on the largest possible DM contribution from the H$\alpha$ emission region obtained by our observations require the presence of the `missing' baryons in the IGM (Equation~\ref{dm_igm}), as expected in the concordance $\Lambda$CDM cosmology.


\acknowledgments

This work was supported by JSPS KAKENHI Grant Number 17J01884, 15J10324, 26800101 and 26287029.
We thank Masaki S. Yamaguchi, Yuki Kikuchi, and Ryou Ohsawa for comments and discussions, and Shriharsh Tendulkar and Cees Bassa for providing us the information of the equatorial coordinates of the host galaxy of FRB~121102 measured on the GMOS broad-band images \citep{ten17}.

This research has made use of the Keck Observatory Archive (KOA), which is operated by the W. M. Keck Observatory and the NASA Exoplanet Science Institute (NExScI), under contract with the National Aeronautics and Space Administration.
The Keck archival data (KOAID=LR.20141119.38797.fits, LR.20141119.40871.fits, LR.20141119.41197.fits, LR.20141119.41528.fits) obtained through the program C220LA (PI: S. Kulkarni) were used.

This research has made use of the Pan-STARRS1 Surveys (PS1) public science archive \citep{cha16,fle16}.
The PS1 and the PS1 public science archive have been made possible through contributions by the Institute for Astronomy, the University of Hawaii, the Pan-STARRS Project Office, the Max-Planck Society and its participating institutes, the Max Planck Institute for Astronomy, Heidelberg and the Max Planck Institute for Extraterrestrial Physics, Garching, The Johns Hopkins University, Durham University, the University of Edinburgh, the Queen's University Belfast, the Harvard-Smithsonian Center for Astrophysics, the Las Cumbres Observatory Global Telescope Network Incorporated, the National Central University of Taiwan, the Space Telescope Science Institute, the National Aeronautics and Space Administration under Grant No. NNX08AR22G issued through the Planetary Science Division of the NASA Science Mission Directorate, the National Science Foundation Grant No. AST-1238877, the University of Maryland, Eotvos Lorand University (ELTE), the Los Alamos National Laboratory, and the Gordon and Betty Moore Foundation.

%


\facility{Subaru (Kyoto~3DII)}
\software{IRAF}


\bibliography{./frb}

\end{document}